\documentclass[prl,twocolumn,superscriptaddress,showpacs,amssymb,amsmath,amsfonts,aps]{revtex4-1}
\usepackage{bm}
\usepackage{graphicx}
\usepackage{dcolumn}
\begin{document}
\title{A precision measurement of the $p$($e,e^\prime p\,$)$\pi^0$ reaction near threshold}

\newcommand*{\ULJUB}{University of Ljubljana, 1000 Ljubljana, Slovenia}
\newcommand*{\SI}{Jo\v{z}ef Stefan Institute, SI-1000 Ljubljana, Slovenia}
\newcommand*{\CAL}{California State University, Los Angeles, Los Angeles, CA 90032}
\newcommand*{\GLAS}{University of Glasgow, Glasgow, G12 8QQ Scotland UK}
\newcommand*{\EDIN}{University of Edinburgh, Edinburgh, EH8 9YL Scotland, UK}
\newcommand*{\STPB}{St.~Petersburg Nuclear Physics Institute, Gatchina, Russia}
\newcommand*{\ORSA}{Institut de Physique Nucleaire, F-91406 Orsay Cedex, France}
\newcommand*{\LUND}{University of Lund, Box 118, SE-221 00 Lund, Sweden}
\newcommand*{\BUDK}{Budker Institute, 630090 Novosibirsk, Russia}
\newcommand*{\RUTG}{Rutgers University, New Brunswick, NJ 08903}
\newcommand*{\SASK}{University of Saskatchewan, Saskatoon, Canada S7N 0W0}
\newcommand*{\YERV}{Yerevan Physics Institute, Yerevan, 0036 Armenia}
\newcommand*{\MIT}{Massachusetts Institute of Technology, Cambridge, MA 02139}
\newcommand*{\KENT}{Kent State University, Kent, OH 44242}
\newcommand*{\UVA}{University of Virginia, Charlottesville, VA 22904}
\newcommand*{\TELAVIV}{Tel Aviv University, Tel Aviv 69978, Israel}
\newcommand*{\NSU}{Norfolk State University, Norfolk, VA 23504}
\newcommand*{\HU}{Hampton University, Hampton, VA 23668}
\newcommand*{\JLAB}{Thomas Jefferson National Accelerator Facility, Newport News, VA 23606}
\newcommand*{\GWU}{The George Washington University, Washington, D.C. 20052}
\newcommand*{\WAM}{College of William and Mary, Williamsburg, VA 23187}
\newcommand*{\INFN}{Istituto Nazionale di Fisica Nucleare, Sezione Sanit\`{a}, I-00161 Rome, Italy}
\newcommand*{\SNU}{Seoul National University, Seoul 151-747, Korea}
\newcommand*{\CMU}{Carnegie Mellon University, Pittsburgh, PA 15213}
\newcommand*{\UMASS}{University of Massachusetts, Amherst, MA 01003}
\newcommand*{\UOK}{University of Kentucky, Lexington, KY 40506}
\newcommand*{\CHMU}{Chiang Mai University, Chiang Mai, Thailand 50200}
\newcommand*{\DU}{Duquesne University, Pittsburgh, PA 15282}
\newcommand*{\SU}{Syracuse University, Syracuse, NY 13244}
\newcommand*{\WEIZ}{The Weizmann Institute of Science, Rehovot 76100, Israel}
\newcommand*{\LBNL}{Lawrence Berkeley National Lab, Berkeley, CA 94720}
\newcommand*{\RIP}{Racah Institute of Physics, Hebrew University of Jerusalem, Jerusalem, Israel 91904}
\newcommand*{\TU}{Temple University, Philadelphia, PA 19122}
\newcommand*{\DUKE}{Duke University, Durham, NC 27708}
\newcommand*{\IDHO}{Idaho State University, Pocatello, ID, 83209}
\newcommand*{\MSU}{Mississipi State University, Starkville, MS 39762}
\newcommand*{\UCM}{Universidad Complutense de Madrid, Madrid, 98040 Spain}
\newcommand*{\XXXX}{UNKNOWN }

\author {K.~Chirapatpimol} \affiliation{\UVA} \affiliation{\CHMU}
\author {M.H.~Shabestari} \affiliation{\UVA}\affiliation{\MSU}
\author {R.A.~Lindgren} \affiliation{\UVA} 
\author {L.C.~Smith} \affiliation{\UVA}
\author {J.R.M.~Annand} \affiliation{\GLAS}
\author {D.W.~Higinbotham} \affiliation{\JLAB}
\author {B.~Moffit} \affiliation{\JLAB}
\author {V.~Nelyubin} \affiliation{\UVA}
\author {B.E.~Norum} \affiliation{\UVA}
\author {K.~Allada} \affiliation{\MIT}
\author {K.~Aniol} \affiliation{\CAL}
\author {K.~Ardashev} \affiliation{\UVA}
\author {D.S.~Armstrong} \affiliation{\WAM}
\author {R.A.~Arndt$^\ddagger$} \affiliation{\GWU}
\author {F.~Benmokhtar} \affiliation{\DU}
\author {A.M.~Bernstein} \affiliation{\MIT}
\author {W.~Bertozzi} \affiliation{\MIT}
\author {W.J.~Briscoe} \affiliation{\GWU}
\author {L.~Bimbot} \affiliation{\ORSA}
\author {A.~Camsonne} \affiliation{\JLAB}
\author {J.-P.~Chen} \affiliation{\JLAB}
\author {S.~Choi} \affiliation{\SNU}
\author {E.~Chudakov} \affiliation{\JLAB}
\author {E.~Cisbani} \affiliation{\INFN}
\author {F.~Cusanno} \affiliation{\INFN}
\author {M.M.~Dalton} \affiliation{\JLAB}
\author {C.~Dutta} \affiliation{\UOK}
\author {K.~Egiyan$^\ddagger$} \affiliation{\YERV}
\author {C.~Fern\`{a}ndez-Ram\`{i}rez} \affiliation{\JLAB}
\author {R.~Feuerbach} \affiliation{\JLAB}
\author {K.G.~Fissum} \affiliation{\LUND}
\author {S.~Frullani} \affiliation{\INFN}
\author {F.~Garibaldi} \affiliation{\INFN}
\author {O.~Gayou} \affiliation{\MIT}
\author {R.~Gilman} \affiliation{\RUTG}
\author {S.~Gilad} \affiliation{\MIT}
\author {J.~Goity} \affiliation{\HU}
\author {J.~Gomez} \affiliation{\JLAB}
\author {B.~Hahn} \ \affiliation{\WAM}
\author {D.~Hamilton} \affiliation{\GLAS}
\author {J.-O.~Hansen} \affiliation{\JLAB}
\author {J.~Huang} \affiliation{\MIT}
\author {R.~Igarashi} \affiliation{\SASK}
\author {D.~Ireland} \affiliation{\GLAS}
\author {C.W.~de~Jager} \affiliation{\JLAB}\affiliation{\UVA}
\author {X.~Jin} \affiliation{\UVA}
\author {X.~Jiang} \affiliation{\RUTG}
\author {T.~Jinasundera} \affiliation{\UVA}
\author {J.~Kellie} \affiliation{\GLAS}
\author {C.E.~Keppel} \affiliation{\HU}
\author {N.~Kolb} \affiliation{\SASK}
\author {J.~LeRose} \affiliation{\JLAB} 
\author {N.~Liyanage} \affiliation{\UVA}
\author {K.~Livingston} \affiliation{\GLAS}
\author {D.~McNulty} \affiliation{\UMASS} \affiliation{\IDHO} 
\author {L.~Mercado} \affiliation{\UMASS}
\author {R.~Michaels} \affiliation{\JLAB}
\author {M.~Mihovilovi\v{c}}\affiliation{\SI} 
\author {S.~Qian} \affiliation{\MIT}
\author {X.~Qian} \affiliation{\DUKE}
\author {S.~Mailyan} \affiliation{\YERV}
\author {V.~Mamyan} \affiliation{\CMU}
\author {S.~Marrone} \affiliation{\INFN}
\author {P.~Monaghan} \affiliation{\MIT}
\author {S.~Nanda} \affiliation{\JLAB}
\author {C.F.~Perdrisat} \affiliation{\WAM}
\author {E.~Piasetzky} \affiliation{\TELAVIV} 
\author {D.~ Protopopescu} \affiliation{\GLAS}
\author {V.~Punjabi} \affiliation{\NSU}
\author {Y.~Qiang} \affiliation{\MIT}
\author {I.A.~Rachek} \affiliation{\BUDK}
\author {A.~Rakhman} \affiliation{\SU}
\author {G.~Ron} \affiliation{\LBNL} \affiliation{\RIP}
\author {G.~Rosner} \affiliation{\GLAS}
\author {A.~Saha$^\ddagger$} \affiliation{\JLAB}
\author {B.~Sawatzky} \affiliation{\TU} \affiliation{\JLAB}
\author {A.~Shahinyan} \affiliation{\YERV}
\author {S.~\v{S}irca} \affiliation{\ULJUB} \affiliation{\SI}
\author {N.~Sparveris} \affiliation{\MIT} \affiliation{\TU}
\author {R.R.~Subedi} \affiliation{\KENT} 
\author {R.~Suleiman} \affiliation{\MIT}
\author {I.~Strakovsky} \affiliation{\GWU}
\author {V.~Sulkosky} \affiliation{\MIT}\affiliation{\JLAB}
\author {J.~Moinelo} \affiliation{\UCM}
\author {H.~Voskanyan} \affiliation{\YERV}
\author {K.~Wang} \affiliation{\UVA}
\author {Y.~Wang} \affiliation{\RUTG}
\author {J.~Watson} \affiliation{\KENT}
\author {D.~Watts} \affiliation{\EDIN}
\author {B.~Wojtsekhowski} \affiliation{\JLAB}
\author {R.L.~Workman} \affiliation{\GWU}
\author {H.~Yao} \affiliation{\TU}
\author {X.~Zhan} \affiliation{\MIT}
\author {Y.~Zhang} \affiliation{\MIT}
\collaboration{The Hall A Collaboration}
\noaffiliation

\date{\today}

\begin{abstract}
New results are reported from a measurement of $\pi^0$ electroproduction near threshold using the $p(e,e^{\prime} p)\pi^0$ reaction.  The experiment was designed to determine precisely the energy dependence of $s-$ and $p-$wave electromagnetic multipoles as a stringent test of the predictions of Chiral Perturbation Theory (ChPT).
The data were taken with an electron beam energy of 1192 MeV using a two-spectrometer setup in Hall A at Jefferson Lab.  For the first time, complete coverage of the $\phi^*_{\pi}$ and $\theta^*_{\pi}$ angles in the $p \pi^0$ center-of-mass was obtained for invariant energies above threshold from 0.5~MeV up to 15~MeV.  The 4-momentum transfer $Q^2$ coverage ranges from 0.05 to 0.155 (GeV/c)$^2$ in fine steps.  A simple phenomenological analysis of our data shows strong disagreement with $p-$wave predictions from ChPT for $Q^2>0.07$ (GeV/c)$^2$, while the $s-$wave predictions are in reasonable agreement.   
\end{abstract}

\maketitle

Neutral pion production from the proton vanishes in the chiral limit of zero quark masses and pion momenta $p_{\pi}\rightarrow 0$.  As a result, the reaction at threshold is particularly sensitive to non-perturbative mechanisms within QCD which break chiral symmetry.  It is also experimentally the most challenging to study.  Pion photo- and electroproduction experiments are now producing data of unprecedented precision to test Chiral Perturbation Theory (ChPT), the low-energy effective field theory of QCD~\cite{bkm95}.  ChPT treats the spontaneous and explicit chiral symmetry breaking in terms of a perturbative expansion in small momenta and quark masses, and makes predictions for the $s-$ and $p-$wave multipoles for the $\gamma N\rightarrow \pi N$ reaction in the near-threshold region.  Within ChPT, the internal structure of the pion and nucleon is systematically parameterized by Low Energy Constants (LEC), while the long-range external $\pi N$ dynamics are fixed by the underlying chiral symmetry. Once the LECs are determined by experiment near threshold, the convergence of
the chiral expansion can be tested by comparing predictions with data taken at energies above threshold. 

Recently $\pi^0$ photoproduction cross-section and polarized photon beam-asymmetry ($\Sigma$) data from the MAMI A2/CB-TAPS experiment~\cite{hor12} were used to test two versions of ChPT.  The relativistic ChPT calculation (RChPT/$\chi$MAID)~\cite{hst13,hlst13,maid} has been carried out to  $\it{O}(p_{\pi}^4)$, while the non-relativistic Heavy Baryon ChPT calculation (HBChPT) is of $\it{O}(p_{\pi}^4)$ for photoproduction (BKM01)~\cite{bkm01} but only of $\it{O}(p_{\pi}^3)$ for $p-$waves in electroproduction (BKM96)~\cite{bkm96}.  Both the BKM01 and RChPT calculations, after fits of LECs to the data, were compatible with the experimental multipoles $E_{0+}, E_{1+}, M_{1+}$ and $M_{1-}$ within an incident photon energy range of 7 to 25 MeV above threshold~\cite{ram12,hst13}.

The pion electroproduction reaction $\gamma^* p\rightarrow p \pi^0$ allows a more stringent test of ChPT, since the four-momentum transfer $Q^2$ and invariant energy $W$ can be varied independently. Chiral $\pi N$ dynamics naturally involve the mass scale $Q^2/m_{\pi}^2$, while the LECs fitted in photoproduction encapsulate higher order processes, involving possibly $N\Delta$ or $\rho,\omega$ degrees of freedom.  The $Q^2$ dependence near threshold may reveal the onset of these short-ranged mechanisms. Until now, only limited kinematic coverage from $\gamma^* p\rightarrow p \pi^0$ threshold experiments is available~\cite{wel92,brk95,dis98,mer02}.  Several older MAMI experiments showed a $Q^2$ dependence of the total cross section near threshold incompatible with HBChPT~\cite{dis98,mer02}, although a new MAMI re-measurement has superceded those data~\cite{mer12}.  The JLAB/Hall A experiment reported here provides the most extensive ($Q^2, W$) coverage of $\pi^0$ electroproduction to date for testing theories of chiral dynamics substantially above threshold.  
 
Under the one-photon-exchange approximation, the $p(e,e^{\prime} p)\pi^0$ cross section 
factorizes as follows:
\begin{equation}
\frac{d\,^3\sigma}{dQ^2 dW d\Omega^*_{\pi}} = J\,\Gamma_v\,\frac{d\sigma}{d\Omega^*_{\pi}}
\end{equation}
where $\Gamma_v$ is the virtual photon flux and the 
Jacobian $J = \partial(Q^2,W)/\partial(E_{e'},\cos\theta_{e'})$ relates the differential
volume element of data binned in $dQ^2 dW$ to the scattered electron kinematics $dE_{e'}\,d\cos\theta_{e'}$.  The $p\pi^0$ center-of-mass (C.M.) differential cross section, $d\sigma/d\Omega^*_{\pi}$, depends 
on the transverse $\epsilon$ and longitudinal $\epsilon_L$  polarization of the virtual 
photon through the response functions $R_T, R_L$ and their interference 
terms $R_{LT}$ and $R_{TT}$:
\begin{eqnarray}
\frac{d\sigma}{d\Omega^*_{\pi}}&=&\frac{p^*_{\pi}}{k_{\gamma}^*}(R_T+\epsilon_L R_L+\epsilon\,R_{TT}\cos 2\phi^*_{\pi} \nonumber \\
&+&\sqrt{2\epsilon_L(\epsilon+1)}\,R_{LT}\cos\phi^*_{\pi}).
\label{eq:2}
\end{eqnarray}
The response functions depend implicitly on $Q^2, W$ and $\theta^*_{\pi}$, the $\pi^0$ C.M. angle, while the angle $\phi^*_{\pi}$ defines the rotation of the $p\pi^0$ plane with respect to the electron scattering plane ($e,e'$).
Other definitions are $\epsilon_L=(Q^2/|k^*|^2)\epsilon$, $\Gamma_v = \alpha E_{e'} k_{\gamma}^* W/2\pi^2 E_e m_p Q^2 (1-\epsilon)$ and $J = \pi W/ E_e E_{e'} m_p$.  Finally $|k^*|$ and $p^*_{\pi}$ are the C.M. momenta of the virtual photon and pion respectively, while $k_{\gamma}^*=(W^2-m_p^2)/2W$ is the real photon equivalent energy.

The $p(e,e^{\prime} p)\pi^0$ experiment was performed in Hall A at Jefferson Lab using the Left High Resolution Spectrometer (LHRS)~\cite{alc04} to detect the scattered electron and the BigBite Spectrometer~\cite{mih12} to detect the coincident proton.  The CEBAF beam was energy-locked to 1192 MeV and delivered to a 6-cm long, 2.54-cm wide cylindrical liquid hydrogen (LH$_2$) target. Beam currents below 5 $\mu$A were used to limit the singles rates in both spectrometers.  Four angular settings for the LHRS ($\theta_{e'}=12.5^\circ,14.5^\circ, 16.5^\circ$ and $20.5^\circ$) covered a nearly continuous $Q^2$ range of $0.05-0.155$ (GeV/c)$^2$ using a 4.4 msr acceptance cut. The LHRS momentum acceptance was centered on the $p\pi^0$ threshold and covered the range $-3\% < \delta p/p <+5\%$.  

\begin{figure}
\begin{center}
\includegraphics[scale=0.3]{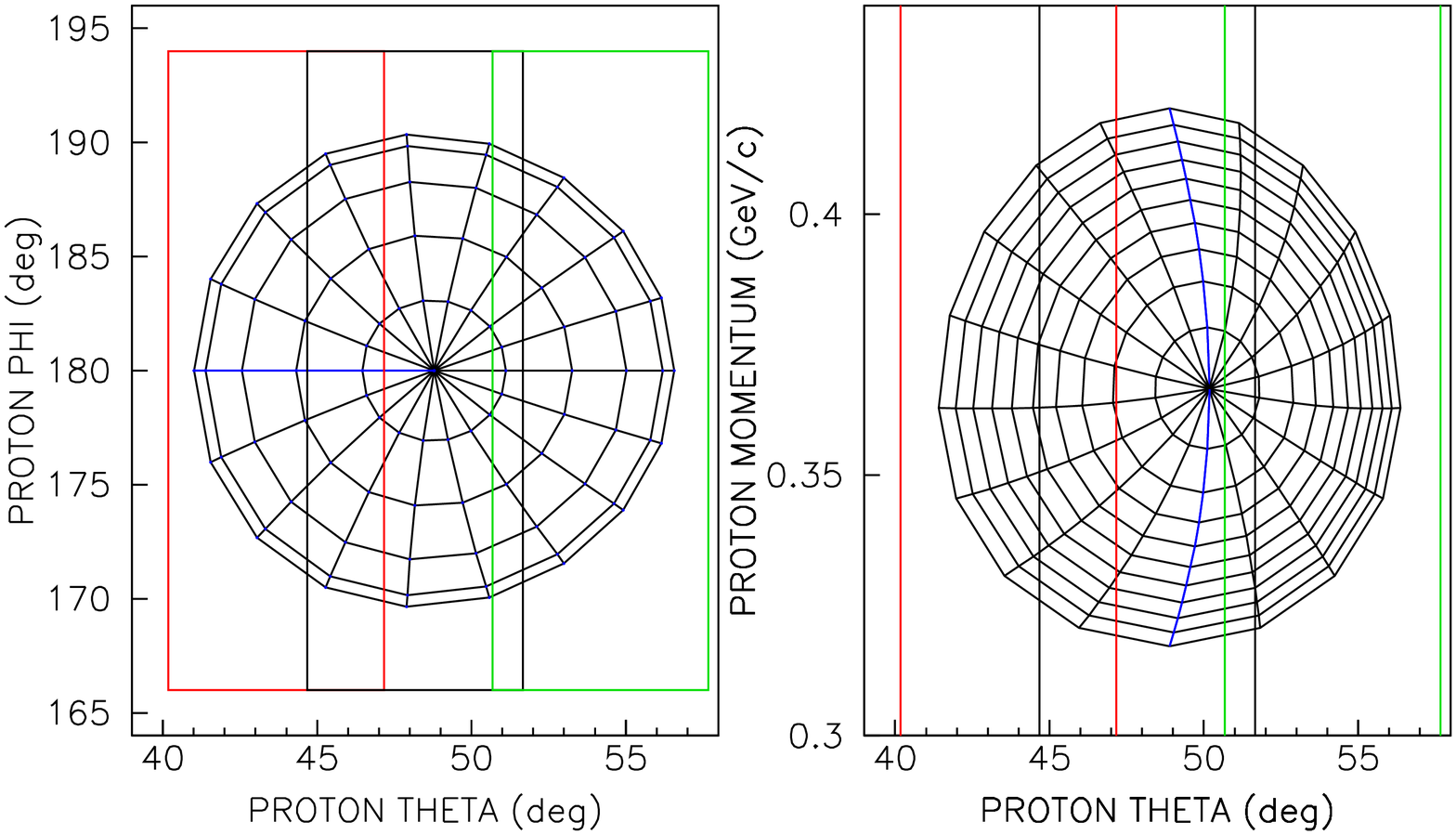}
\caption{Left: Overlap between three BigBite Spectrometer proton laboratory angle settings (colored boxes) and  $p\pi^0$ center-of-mass bins at $Q^2$=0.135 (GeV/c)$^2$ and $\Delta W$=9.5 MeV. Radial and concentric lines separate bins of $\phi^*_{\pi}$ and $\theta^*_{\pi}$, respectively. Only 5 out of 9 $\theta^*_{\pi}$ bins are shown. The blue line shows $\phi^*_{\pi}$=180.  Right: Radial and concentric lines separate bins of $\theta^*_{\pi}$ and $\Delta W$, respectively,
projected onto proton lab momentum $p_p$ and $\theta_p$.  Bins to the (left,right) of the blue line correspond to ($\phi^*_{\pi}$=180$^\circ$, $\phi^*_{\pi}$=0$^\circ$). The innermost circle represents $\Delta W$ = 0.5 MeV.}
\label{accept}
\end{center}
\end{figure}

Three angular settings of the BigBite were used ($\theta_p$=43.5$^\circ$,48$^\circ$ and 54$^\circ$) which provided full coverage (Fig.~\ref{accept}) of the proton cone up to an invariant energy above threshold of $\Delta W$=15 MeV (at the largest $Q^2$).  The BigBite momentum acceptance covered the range (0.25$<p_p<$0.5 GeV/c), limited by the target energy loss at low momentum and the thresholds on the $E-\Delta E$ scintillator counters at high momentum. The low momentum cutoff was achieved using a thin (25 $\mu$m) Ti exit window in the target scattering chamber and a helium bag for transport up to and between the BigBite drift chambers.  Absolute normalization, energy and angle calibrations in both spectrometers were checked at each kinematic setting using elastic scattering runs with LH$_2$ and thin solid targets.  

\begin{figure}[t]
\begin{center}
\includegraphics[width=0.47\textwidth]{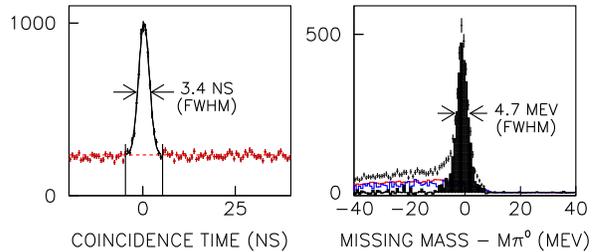}
\caption{Left: Coincidence timing between the LHRS and BigBite. Events belonging to the true coincidence peak were selected using cuts indicated by the vertical lines, while random coincidences were selected from the region highlighted in red.  Right: Missing mass distribution at $Q^2$=0.15 (GeV/c)$^2$ for the invariant mass range $0<\Delta W<10$~MeV.  Background events from random coincidences (red) and target cell windows (blue) were subtracted from the raw distribution, leaving the $\pi^0$ missing mass peak shown in gray.}
\label{timing}
\end{center}
\end{figure}

Scintillator hodoscopes provided the primary triggers for both spectrometers.  A gas threshold \v{C}erenkov detector in the LHRS provided electron identification with 99$\%$ efficiency.  Signals from either $E$ or $\Delta E$ scintillator planes at the rear of BigBite were used in the coincidence trigger, while signal thresholds in both the hodoscopes and multi-wire drift chambers were set to suppress minimum ionizing tracks from pions.  Final proton identification was made using $E-\Delta E$ cuts on the highly segmented scintillators.  The path-length corrected coincidence time distribution between the LHRS and BigBite is shown in Fig.~\ref{timing}.  A 10 ns wide cut centered on the peak was used to select true coincidences, while a 30 ns cut (excluding the peak) selected random coincidences for subtraction.  Selection of the $p\pi^0$ final state required calculation of the missing-mass $M$ after reconstruction of the detected particle's 3-momenta:
\begin{equation}
M^2=(E+m_p-E_{e'}-E_p)^2-(\vec{p}_e-\vec{p}_{e'}-\vec{p}_p)^2
\end{equation}
The experimental missing-mass distribution is also shown in Fig.~\ref{timing} before and after subtraction of both random coincidences and target-window contributions.  The latter background was estimated using cuts on $\Delta W$ below the $\pi^0$ threshold.

Before binning the data, both incident and scattered electron energies were corrected for ionization losses in the LH$_2$ and target windows, using the calculated entrance and exit paths with respect to the measured target interaction vertex. Proton transport energy losses through the target, Ti window and BigBite were also corrected for each event.  Acceptance corrections were derived from a Monte-Carlo simulation of both spectrometers, using the Dubna-Mainz-Taipei (DMT) model~\cite{dmt99} as a physics event generator. Special care was taken to incorporate into the simulation radiative correction and straggling losses, a fine-mesh magnetic field map for the BigBite, and the measured energy and angular resolution and energy calibration determined from elastic scattering runs, in order to properly account for their systematic effects near threshold. The dominant sources of systematic uncertainty are target window background subtraction, accidental coincidence corrections and LHRS central momentum calibration, which combined contribute to the overall normalization error of 20\% near threshold at low $Q^2$ decreasing to 7\% for data above threshold at higher $Q^2$.   

\begin{figure}[t]
\begin{center}
\includegraphics[width=0.51\textwidth]{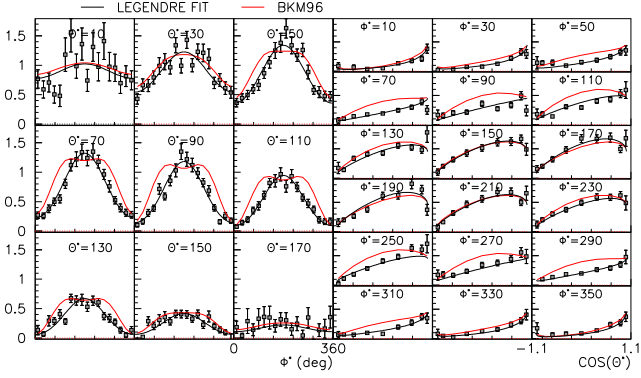}
\caption{Differential cross sections for $p(e,e^{\prime}p)\pi^0$ from this experiment at $Q^2$ = 0.135 (GeV/c)$^2$ and $\Delta W$ = 9.5 MeV binned in $p\pi^0$ center-of-mass angles $\phi^*_{\pi}$ and and cos $\theta^*_{\pi}$.  See text for description of curves.  Units are $\mu b/sr$. Errors are statistical only.}
\label{diffx}
\end{center}
\end{figure}

Events were accumulated using (12,30,18,9) bins for ($Q^2, \Delta W, \phi^*_{\pi}, \theta^*_{\pi}$) respectively, with a cut of $\pm$10 MeV on the missing-mass peak. The $\Delta W$ bin width was 1 MeV and the LHRS acceptance extended up to $\Delta W$=30 MeV, although with reduced C.M. coverage.  The average $Q^2$ bin width was 0.01 (GeV/c)$^2$. Figure~\ref{diffx} shows typical differential cross sections for each $\phi^*_{\pi}$ and $\theta^*_{\pi}$ bin obtained at $Q^2$=0.135 (GeV/c)$^2$ and $\Delta W$ = 9.5 MeV.  The curve labeled BKM96 is the HBChPT prediction from Bernard~\textit{et al.}~\cite{bkm96}, which uses LECs fitted to older photoproduction data from MAMI and electroproduction data at $Q^2$=0.1 (GeV/c)$^2$ from MAMI and NIKHEF.  The other curve is an empirical fit to the data which we use to obtain the total cross section $\sigma_{tot}$.  The empirical fit uses the form in Eq.(2) and expands the response functions with Legendre polynomials $P_l(x)$, where $x=\cos \theta^*_{\pi}$,
\label{eq3}
\begin{eqnarray}
R_T+\epsilon_L R_L &=& A^{T+L}_{0}+A^{T+L}_{1}\,P_1(x)+A^{T+L}_{2}\,P_2(x) \label{eq:3a}\\
R_{TT} &=& A^{TT}_0\,(1-x^2) \label{eq:3b}\\
R_{LT} &=& (A^{LT}_0+A^{LT}_1\,P_1(x))\,(1-x^2)^{1/2}. \label{eq:3c}
\end{eqnarray}
The total cross section $\sigma_{tot}$ is given by $4\pi\frac{p^*_{\pi}}{k_{\gamma}^*}\,A^{T+L}_0$. 

\begin{figure}[t]
\begin{center}
\includegraphics[width=0.53\textwidth]{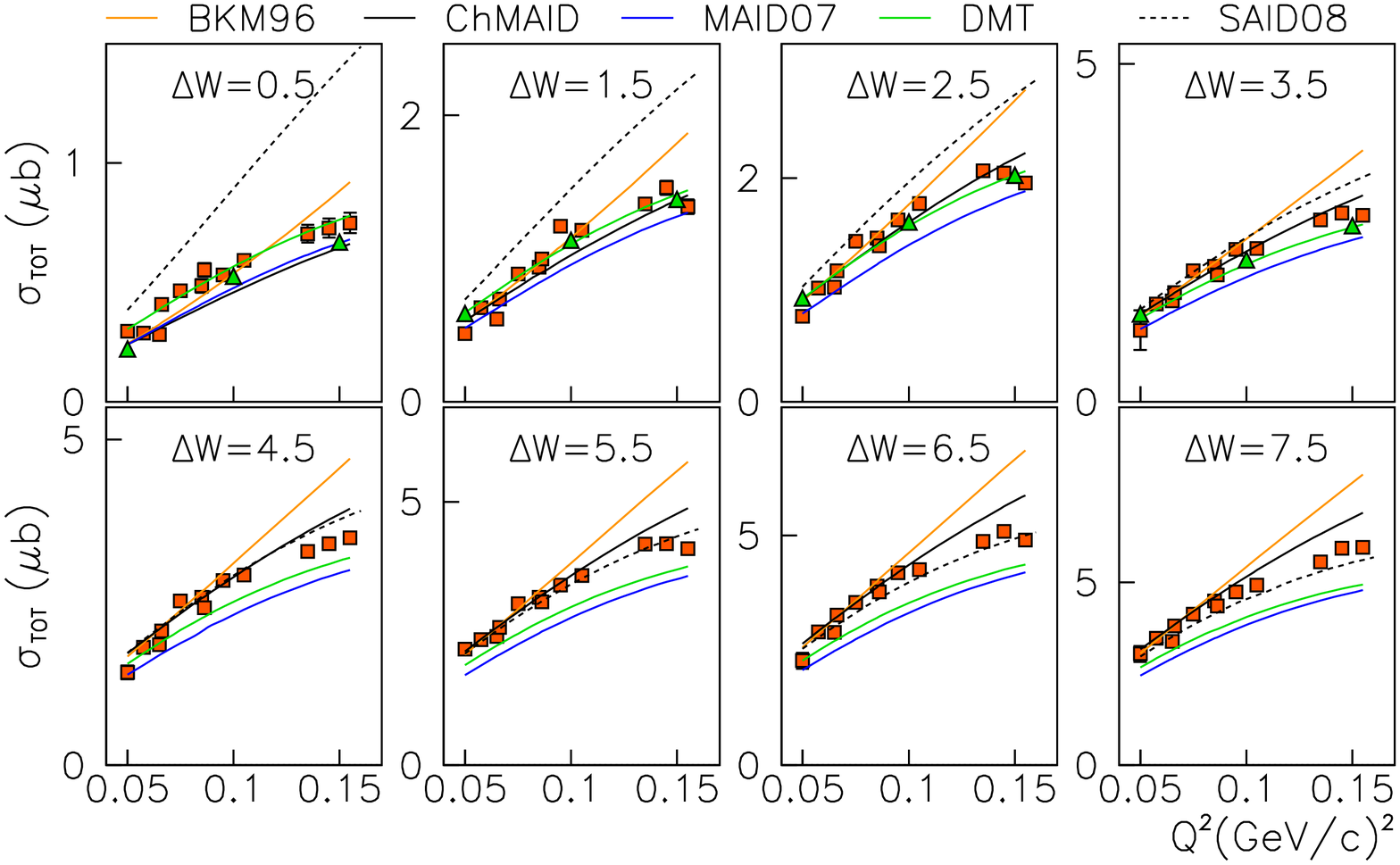}
\caption{Total cross section for $p(e,e^{\prime}p)\pi^0$ as a function of $Q^2$ for different bins in $\Delta W$ (invariant mass above threshold) for ($\Box$) this experiment and ($\triangle$) MAMI~\cite{mer12}.  Units of $\Delta W$ are MeV. Errors are statistical only.}
\label{totcrs}
\end{center}
\end{figure}

The $Q^2$ dependence of $\sigma_{tot}$ is shown in  Fig.~\ref{totcrs} for different $\Delta W$ bins starting 0.5 MeV above threshold. Two ChPT calculations are shown (BKM96~\cite{bkm96}, $\chi$MAID~\cite{hlst13}), along with the SAID08 solution~\cite{said} and phenomenological models (DMT~\cite{dmt99}, MAID~\cite{maid07}) which have been fitted to the world data on pion photo- and electroproduction.  Compared to the linear $Q^2$ dependence of the HBChPT/BKM96 curve, our $\sigma_{tot}$ measurement shows a bending over at higher $Q^2$ similar to the phenomenological models and the RChPT/$\chi$MAID theory. At lower $Q^2$, both ChPT calculations are consistent with our data over the entire $\Delta W$ range shown here.  Note that two of the RChPT LECs were fitted to a new MAMI re-measurement~\cite{mer12} (triangles in Fig.~\ref{totcrs}) of earlier $Q^2>0$ experiments, while the remaining LECs were fitted to the $Q^2=0$ A2/CB-TAPS data~\cite{hor12}. 

\begin{figure}[t]
\begin{center}
\includegraphics[width=0.52\textwidth]{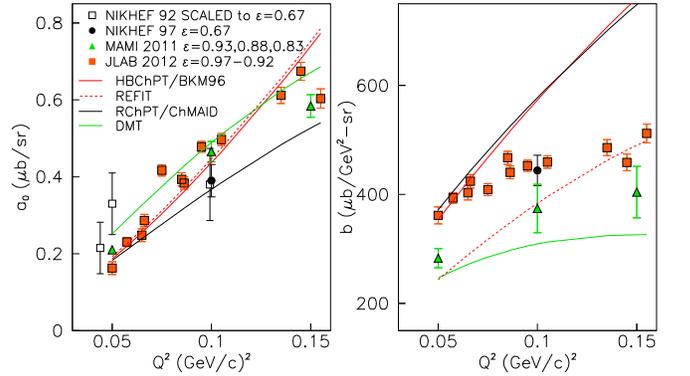}
\end{center}
\caption{The $Q^2$ dependence of $a_0$ (left) and $b$ (right) from the fits of Eq. 7 to the Legendre coefficient $A^{T+L}_{0}$. The theory curves are calculated for the beam energy of our experiment (1192 MeV). For the curve labeled REFIT the BKM96 LEC $b_P$ has been lowered from 13 to 9.3~(GeV)$^{-3}$ (see text and Fig.~6).  Errors are statistical only.}
\label{abplot}
\end{figure}

Near threshold, the $s-$ and $p-$wave decomposition of $\sigma_{tot}$ can be obtained by fitting the $p^*_{\pi}$ dependence of $A^{T+L}_{0}$ according to
\begin{equation}
A^{T+L}_{0}=a_0 + b\, |p^*_{\pi}|^2.
\end{equation}
The $b$ coefficient parameterizes the contribution of $p-$wave multipoles arising from their intrinsic $p^*_{\pi}$ dependence near threshold, while $a_0$ fits the combination $|E_{0+}|^2 + \epsilon_{L}|L_{0+}|^2$ of $s-$wave multipoles extrapolated to threshold.  The $L_{0+}$ multipole dominates $a_0$ over our $Q^2$ range due to a large $\epsilon_L$ factor.  The extraction of $a_0$ and $b$ from fitting our data up to $\Delta W=9.5$~MeV is shown in Fig.~\ref{abplot}, along with fits to the newest MAMI data~\cite{mer12} up to $\Delta W=3.5$~MeV (the limit of their measurement) and previous results from NIKHEF~\cite{wel92,brk95}.  There is good agreement of both $a_0$ and $b$ with the chiral model predictions for our lowest $Q^2$ points.  For higher $Q^2$, the HBChPT curve describes $a_0$ better than RChPT. However the strong disagreement of our $b$ coefficient with both chiral curves for $Q^2>0.07$~GeV$^2$ suggests at least one of the $p-$wave multipoles is described incorrectly in the calculations.  The $Q^2$ dependence of $b$ from fitting the MAMI data is qualitatively similar, although with larger errors, due to the smaller $\Delta W$ range of their data. 

\begin{figure}[t]
\begin{center}
\includegraphics[width=0.51\textwidth]{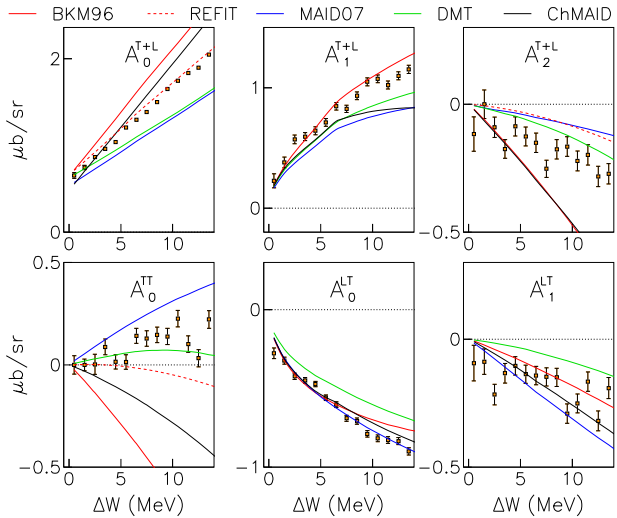}
\end{center}
\caption{The $\Delta W$ dependence of Legendre coefficients from fits to our data at $Q^2$ = 0.135 (GeV/c)$^2$.   Note that the $\Delta W$=9.5~MeV bin corresponds to the Legendre fit shown in Fig.~\ref{diffx}. Errors are statistical only.}
\label{legen}
\end{figure}

Further insight can be obtained from the $\Delta W$ dependence of the Legendre coefficients in the $Q^2 > 0.07$~(GeV/c)$^2$ region.  This is shown in Fig.~\ref{legen} at $Q^2$=0.135~(GeV/c)$^2$.  While all models are in good agreement with our data near threshold, the theory curves for $A^{T+L}_{0}$, $A^{T+L}_{2}$ and $A^{TT}_{0}$ show large variations above $\Delta W$=3~MeV.  These coefficients are particularly sensitive to the $p-$wave multipole combinations $P_3=2M_{1+}+M_{1-}$ and $P_2=3E_{1+}-M_{1+}+M_{1-}$, while $A^{TT}_{0}$ is also sensitive to the combination $\Delta P_{23}^2=(P_2^2-P_3^2)/2$.  Our fit result for $A^{TT}_{0}$ is close to zero over the $\Delta W$ range of our data, which implies $P_2^2 \approx P_3^2$ or $M_{1+}/M_{1-}\approx-2$ (neglecting the weak electric quadrupole $E_{1+}$).
 
Only the DMT model predicts $A^{TT}_{0}\approx 0$ for $\Delta W<15$~MeV, largely due to their calculation of the $M_{1-}$ multipole~\cite{kam01}, the value of which is substantially larger than predicted by ChPT.  A similar result was obtained from disperson relations~\cite{kam02}. In the BKM96 theory, which uses a $\it{O}(p^3)$ $p-$wave expansion, it is not possible to separately adjust $M_{1+}$ and $M_{1-}$, since only $P_3$ is controlled by a single LEC $b_P$. By reducing $b_P$ in the calculation from 13.0 to 9.3~(GeV)$^{-3}$, we can improve agreement with both $A^{T+L}_{0}$ and $A^{TT}_{0}$ as shown in Fig.~\ref{legen} by the curve labeled REFIT.  However this adjustment worsens the agreement with $p-$waves at lower $Q^2$, as indicated by the REFIT $b$ curve in Fig.~\ref{abplot}.  Moreover, a different adjustment of $b_P$ is required to match our measurement of $A^{T+L}_{2}$.  

The $\it{O}(p^4)$ RChPT calculation~\cite{hlst13} predicts a nearly identical $Q^2$ dependence for the $b$ curve in Fig.~\ref{abplot} as the $\it{O}(p^3)$ HBChPT theory.  At leading-order and next-to-leading order, $P_3$ is controlled by a single $\it{O}(p^3)$ LEC $d_9$, similarly to HBChPT~\cite{hst13}. However $d_9$ is highly constrained by the $Q^2$=0 photoproduction fits, and there is almost no room for adjustment.  
Other $\it{O}(p^4)$ LECs, which explicitly control $Q^2$ dependent terms, either do not appreciably affect the $p-$wave multipoles, or effect the same $Q^2$ response as $b_P$.


Despite the very different LEC composition of HBChPT and RChPT, it appears neither calculation can be adjusted to 
agree with the $Q^2$ trend of our $p-$wave data.  Furthermore this discrepancy occurs well within the $\Delta W$ range where photoproduction $p-$waves are well described at $\it{O}(p^4)$~\cite{ram12}.  Our data therefore suggest that higher powers of $Q^2$ are needed in the ChPT formalism, while the onset of disagreement ($Q^2>$0.07) implies a $t-$channel energy scale above the pion mass.  Similar discrepancies in ChPT calculations of nucleon form factors were removed by including vector mesons as dynamical degrees of freedom~\cite{kub01}.  Our data could provide strong constraints to analogous extensions of pion electroproduction calculations.

In summary, a JLAB/Hall A experiment has measured for the first time both the $Q^2$ and extended $\Delta W$ dependence of the threshold $p(e,e^{\prime} p)\pi^0$ reaction with full C.M. coverage and fine binning.  Our phenomenological fit of the data shows reasonable agreement with two leading ChPT theories for $s-$waves, while chiral predictions of $p-$wave contributions strongly diverge from our data for $Q^2> 0.07$~(GeV/c)$^2$.  We use a Legendre decomposition of our cross sections to show there is insufficient flexibility in the low energy constants available for $p-$waves to account for the $Q^2$ discrepancy. 

\begin{acknowledgments}

The collaboration wishes to acknowledge the Hall A technical staff and the Jefferson Lab 
Accelerator Division for their support.  Special thanks goes to Ulf-G.~Mei$\beta$ner for his support in the early phases of this experiment. This work was supported by the U.S. Department of 
Energy and the U.S. National Science Foundation and in particular DOE contract DE-FG02-97ER41025
and NSF MRI Award No. 021635. Jefferson Science Associates operates 
the Thomas Jefferson National Accelerator Facility under DOE contract DE-AC05-06OR23177.

$^{\ddagger}$ Deceased.

\end{acknowledgments}


\end{document}